\documentclass{PoS}

\usepackage{amssymb,amsmath,amsbsy}
\usepackage{mathrsfs}
\usepackage{mathtools}
\usepackage{float}

\newcommand{\newc}{\newcommand}
\newc{\be}{\begin{equation}}
\newc{\ee}{\end{equation}}
\newc{\bal}{\begin{align}}
\newc{\eal}{\end{align}}
\newc{\ba}{\begin{eqnarray}}
\newc{\ea}{\end{eqnarray}}
\newc{\bea}{\begin{eqnarray*}}
\newc{\eea}{\end{eqnarray*}}
\newc{\D}{\partial}
\newc{\shh}{\lambda_{h}}
\newc{\shf}{\lambda_{h\phi}}
\newc{\sff}{\lambda_{\phi}}
\newc{\sss}{\lambda_{s}}
\newc{\ssisi}{\lambda_{\sigma}}
\newc{\sssi}{\lambda_{s \sigma}}
\newc{\shs}{\lambda_{h s}}
\newc{\shsi}{\lambda_{h \sigma}}
\newc{\sfs}{\lambda_{\phi s}}
\newc{\sfsi}{\lambda_{\phi \sigma}}
\newc{\shz}{\lambda_{h}^{(0)}}
\newc{\shfz}{\lambda_{h\phi}^{(0)}}
\newc{\sfiz}{\lambda_{\phi}^{(0)}}
\newc{\sssz}{\lambda_{s}^{(0)}}
\newc{\ssfz}{\lambda_{s\phi}^{(0)}}
\newc{\som}{\sin\omega}
\newc{\com}{\cos\omega}
\newc{\sth}{\sin\theta}
\newc{\cth}{\cos\theta}
\newc{\stom}{\sin^2\omega}
\newc{\ctom}{\cos^2\omega}
\newc{\stth}{\sin^2\theta}
\newc{\ctth}{\cos^2\theta}
\newc{\ie}{{\it i.e.} }
\newc{\eg}{{\it e.g.} }
\newc{\etc}{{\it etc.} }
\newc{\etal}{{\it et al.}}

\newcommand{\lapproxeq}{\lower .7ex\hbox{$\;\stackrel{\textstyle
<}{\sim}\;$}}
\newcommand{\gapproxeq}{\lower .7ex\hbox{$\;\stackrel{\textstyle
>}{\sim}\;$}}
\newcommand{\stackdown}[2]{\lower 1.4ex\hbox{$\;\stackrel{\textstyle{#1}}
{\scriptstyle{#2}}\;$}}

\usepackage{subcaption}
\usepackage{slashed}

\title{ \centering A Fermionic bi-Doublet Effective Field Theory for Dark Matter}

\ShortTitle{A Fermionic bi-Doublet EFT for DM}

\author{\speaker{D. Karamitros}\\
        University of Ioannina, Greece\\
        E-mail: \email{dkaramit@cc.uoi.gr}}

\abstract{We study  an effective field theory
which includes the  Standard Model extended by a Dark Sector consisting of
two  fermionic $SU(2)_{L}$-doublets. A $Z_2$ parity  guarantees that, after electroweak symmetry breaking, 
the lightest neutral particle is stable, acting as a WIMP.
The dark sector interacts with the Higgs and gauge bosons
through renormalizable and non-renormalizable $d=5$ operators.
We find that a WIMP with a mass around the electroweak scale,
\textit{i.e.} accessible  at the LHC,
is consistent with collider and astrophysical data only when
non-trivial magnetic dipole interactions with the gauge bosons exist.
}

\FullConference{Corfu Summer Institute 2016 "School and Workshops on Elementary Particle Physics and Gravity"\\
		31 August - 23 September, 2016\\
		Corfu, Greece}

\begin{document}

\section{Introduction}
\label{sec:intro}

There is a number of evidence suggesting  that the mass content of the universe is dominated by Dark Matter (DM). 
From CMB measurements, the DM seems to account for about 25\% of the total energy of the universe\cite{Ade:2015xua}. 
One of the  most promising DM candidates is the so-called Weakly Interacting Massive Particle (WIMP). The WIMP is assumed to be a massive
electrically neutral stable particle interacting weakly with the Standard Model (SM). Under these assumptions, the mass of the WIMP seems to lie
naturally at the electroweak scale, due to the so-called \textit{WIMP-miracle}~\cite{Hooper:2009zm}. This makes the WIMP accessible at LHC as well as direct detection experiments.   

Here we present the work of~\cite{Dedes:2016odh}, where the SM is extended by a pair of fermionic $SU(2)_{L}$-doublets, which constitutes the Dark Sector (DS). Assuming that the SM and the DS do not mix, due to a $Z_2$ parity, the complete set of $d=5$ non-renormalizable operators is introduced. We show that after electroweak (EW) symmetry breaking there is a stable neutral particle, which can act as a WIMP.
The EFT at hand generalises the discussion on the bi-doublet DM scenario. Among the  models incorporating a pair of fermion doublets in their low-energy spectrum, one finds the \textit{higgsino} DM case~\cite{Roszkowski:1991ng}, some simplified models such as the doublet-triplet DM~\cite{Dedes:2014hga}, non-supersymmetric SO(10) GUTs~\cite{Nagata:2015dma} and its left-right symmetric subgroup~\cite{Garcia-Cely:2015quu}. 

Performing a phenomenological analysis, we show that a viable WIMP with a mass close to the EW scale, i.e. suitable for LHC searches,
acquires sizeable magnetic dipole moments with the gauge bosons.

\section{The EFT content}
\label{sec:EFT_cont}
In the SM we add  a pair of fermion $SU(2)_L$-doublets, $D_{1,2}$,  with opposite hypercharges, $Y(D_{1})=-Y(D_{2})=-1$. We impose a $Z_2$ parity 
which separates the DS from the SM and ensures that the lightest neutral particle is stable and thus a WIMP candidate. Apart from the renormalizable
interactions, we also introduce the complete set of $d=5$ non-renormalizable operators, which are responsible for the dipole 
interactions between the DS fermions and the SM gauge bosons, the Yukawa interactions and the mass splitting between the components of the doublets. 
 
\subsection{The Yukawa interactions}
Since there are no renormalizable interactions between the Higgs boson and $D_{1,2}$, they appear at the $d=5$ level. The $d=5$ Yukawa along with the mass terms are\footnote{The spinor and gauge indices are suppressed for simplicity.}   

\begin{eqnarray}
   & -\mathcal{L}_{\rm mass + Yukawa} \   \supset  \
    \ \frac{y_{1}}{2\, \Lambda } \: (   H^{T} \epsilon D_{1} )\: (   \, H^{T} \epsilon D_{1} )
   \ + \ \frac{y_{2}}{2 \Lambda } \: (H^{\dagger } D_{2} )\: (H^{\dagger } D_{2} )
   \label{L_Yuk} \\[2mm]
   & \ + \ \frac{y_{12}}{\Lambda } \:  (   H^{T} \epsilon D_{1} )\: (H^{\dagger } D_{2} ) 
   \ + \ \frac{\xi_{12}}{\Lambda} \: ( \, D_{1}^{T} \epsilon D_{2} ) (H^{\dagger } H_{}) 
   \ + M_{D}\:   {D}_{1}^{T} \epsilon {D}_{2} \nonumber+ \ \mathrm{H.c.},
\end{eqnarray}
where $\Lambda$ is the cut-off of the EFT and $\epsilon$ is the $SU(2)_L$ anti-symmetric tensor (in the fundamental representation). Also, for simplicity we assume that the parameters are real numbers, while the mass parameter $M_D$ can be redefined to be positive. Finally, as it can be seen from eq.(\ref{L_Yuk}), there are four independent operators with their respective Wilson coefficients $y_{1,2,12}$ and $\xi_{12}$.

\subsection{The dipole interactions}

Apart from the operators of eq.~\ref{L_Yuk}, there are also interactions with the gauge bosons at $d=5$ level. These are

\begin{eqnarray}
  \mathcal{L}_{\rm dipoles}  \supset & \frac{d_{\gamma}}{\Lambda}\:  D_{1}^{T} \: \sigma^{\mu\nu}  \: \epsilon D_{2}\: 
  B_{\mu\nu}   \  + \    \frac{d_{W}}{\Lambda}\:  \left( D_{1}^{T} \: \sigma^{\mu\nu} \: \epsilon \vec{\tau} \: D_{2} \right) \cdot \: 
  \vec{W}_{\mu\nu} \ + \ 
  \nonumber \\[2mm]
  & \frac{i\, e_{\gamma}}{\Lambda}\:  D_{1}^{T} \: \sigma^{\mu\nu}  \: \epsilon D_{2}\: 
  \widetilde{B}_{\mu\nu}  \ + \
  \frac{i\, e_{W}}{\Lambda}\:  \left( D_{1}^{T} \: \sigma^{\mu\nu} \: \epsilon \vec{\tau} \: D_{2} \right) \cdot\: 
  \vec{\widetilde{W}}_{\mu\nu} \ + \  
  \mathrm{H.c.}\;,  
  \label{dipoles}
 \end{eqnarray} 
$B_{\mu \nu}$ ($\vec{W}_{\mu \nu}$) the $U(1)_Y$ ($SU(2)_L$) gauge boson, $d_{\gamma}$ and $d_W$ are real numbers. Additionally, since we are not concerned about $CP$ violation, $e_{\gamma}=e_{W}=0$.

\subsection{Symmetries of the Dark sector}

\subsubsection*{The custodial symmetry}
It is  known~\cite{Sikivie:1980hm} that, the SM Higgs sector is invariant under 
a global $SU(2)_R$ (custodial) symmetry. defining

\begin{equation}
\mathcal{H}=\left(    \begin{array}{ll}
               -H^{0*} & H^{+}  \\ H^{-} & H^{0}
              \end{array}           \right),
\label{H_cust}
\end{equation}
 the the SM Higgs sector is invariant under $SU(2)_{L}\times SU(2)_{R}$ with the transformation rule $ \mathcal{H} \to U_{L} \mathcal{H} U_{R}$.  It turns out that in the EFT at hand the Yukawa sector exhibits the same symmetry, when 
\begin{equation}
y_{1}=y_{2}=-y_{12}.
\label{y_cust}
\end{equation}
This can be seen by defining $y \equiv - y_{12}$ and
 
\begin{equation}
\mathcal{D}=\left(    \begin{array}{ll}
               D_{1}^{0} & D_{2}^{+}  \\ D_{1}^{-} & D_{2}^{0}
              \end{array}           \right),
\label{D_cust}
\end{equation}
which transforms as $ \mathcal{D} \to U_{L} \mathcal{D} U_{R}$. Then the equation ~\ref{L_Yuk} obtains the form:

\begin{equation}
-\mathcal{L}_{Yuk} \supset  \frac{y}{\Lambda}\:  \left[ Tr (\mathcal{H}^{\dagger}\mathcal{D} )\right]^2 + M_D \: det\mathcal{D}+H.c.,
\end{equation} 
which is clearly invariant under  $SU(2)_{L} \times SU(2)_{R}$. 

\subsubsection*{The charge conjugation symmetry}
In addition to the custodial symmetry, there is also a \textit{charge conjugation} (c.c.) symmetry, which is a symmetry of the entire set of $d=5$ operators. For $y_{1}=y_{2}=y$ (and $\forall y_{12}$), the interactions~\ref{L_Yuk} and~\ref{dipoles} are invariant under exchanging $D_{1} \to D_{2}$ according to\footnote{In general the Higgs field is  transformed as $H \to H^{\dagger}$, but in the Kibble parametrization $H$ remains unaffected.}  

\begin{equation}
C^{-1} D_{2a} C = \epsilon^{ba}D_{1b} .
\end{equation} 
This symmetry, basically, exchanges the columns of the matrix \ref{D_cust}.

Finally, we should point out that in our phenomenological analysis we are going to study two \textit{Benchmark} scenarios in the  the c.c. symmetric limit i.e. $y_{1}=y_{2}$. These two case are:

\begin{equation}
\mathrm{(a)}\ y_{12}=-y, \: \mathrm{(b)}\ y_{12}=0.
\label{Benchmark}
\end{equation}  
The first one is the $SU(2)_{R}$ symmetric limit, while the second violates the custodial symmetry, but is employed since it gives us a distinct mass spectrum. 
%

\section{The physical states}
\label{sec:physical}

\subsection{Mass spectrum}

After EW symmetry breaking, the Higgs field is shifted by its vacuum expectation value (vev), $v$, 
resulting to  mixing between the components of $D_{1,2}$. After rotating to the mass basis,  physical states are

 \begin{eqnarray}
   \chi_{1}^{0} = \frac{1}{\sqrt{2}} \: ( D_{1}^{0} + D_{2}^{0} ) \;, \quad &
  \chi_{2}^{0} = -\frac{i}{\sqrt{2}} \: ( D_{1}^{0} - D_{2}^{0} ) \;,  \nonumber \\[1mm]
  \chi^{+} = i \: D_{2}^{+} \;, \qquad  &\chi^{-} = i \: D_{1}^{-} \;.
  \label{rotation}
 \end{eqnarray}
For these particles the mass terms become 
\begin{equation}
\mathcal{L}\supset - m_{\chi^{\pm}} \chi^{-}\chi^{+} - \frac{1}{2} \sum\limits_{i=1}^2  m_{\chi^{0}_{i}} \chi^{0}_{i}\chi^{0}_{i}+H.c. \, ,
\label{masses}
\end{equation}
where the masses are
  \begin{eqnarray}
  & m_{\chi^{\pm}}  =  M_{D}\ + \ \xi_{12} \: \omega \;, \nonumber 
   \\[1mm]
  & m_{\chi_{1}^{0}} =  m_{\chi^{\pm}} \ + \ \omega\:  (y - y_{12}) \;,  \qquad \omega \equiv \frac{v^{2}}{\Lambda} \;, \label{masses}
   \\[1mm]
  & m_{\chi_{2}^{0}}  =  m_{\chi^{\pm}}\ - \ \omega \: (y + y_{12}) \;.  \nonumber 
 \end{eqnarray}
As stated in the previous section, we consider the two Benchmark scenarios shown in \ref{Benchmark}.
These produce two distinct hierarchies for the masses $ m_{\chi^{\pm}}, \, m_{\chi^{0}_{1,2}}$.
 \begin{figure}
\begin{center} 
\includegraphics[width=10cm]{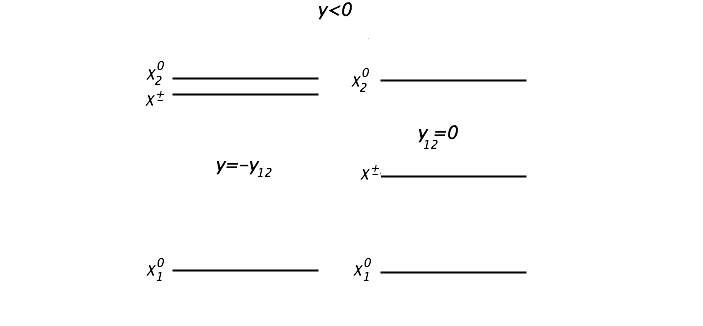}  
\caption{The mass hierarchies  for the two Benchmark scenarios under study.}
\label{hierarchies} 
\end{center}
 \end{figure}

The hierarchies are shown in Fig.~\ref{hierarchies}. 
We note that for $y<0$ the lightest particle is always $\chi_{1}^{0}$, while for  $y>0$, the lightest particle becomes $\chi_{2}^{0}$ without changing the phenomenology. Therefore, for the following analysis we are going to restrict $y$ to be negative, which makes $\chi_{1}^{0}$ our WIMP candidate.    
The masses of the particles for the two cases under study are:  

\noindent
\textbf{(a)} $y=-y_{12}<0$. In this case, the heavy fermion is degenerate with the charged one, where the various masses are given by   
\[  m_{\chi^{0}_{2}}=m_{\chi^{\pm}}, \, m_{\chi^{0}_{1}} = m_{\chi^{\pm}}-2 \omega |y|. \]
\textbf{(b)} $y<0, \, y_{12}=0$. There is no degeneracy between the particles and the various masses are  
\[ m_{\chi^{0}_{2}} = m_{\chi^{\pm}}+2 \omega |y| , \, m_{\chi^{0}_{1}} = m_{\chi^{\pm}}-2 \omega |y|. \]

\subsection{Interactions}

We calculate the Yukawa interactions for the rotated fields. The Lagrangian describing the 3-point DS-Higgs interactions\footnote{The 4-point interactions are not shown for simplicity.} is

\begin{eqnarray}
  \mathcal{L}_{\rm \chi \, \chi \,h}^{\rm dim=5} \supset & - Y^{h\chi^{-} \chi^{+}} \: h \: \chi^{-} \: \chi^{+}
  \  - \ \frac{1}{2}\: Y^{h\chi_{i}^{0} \chi_{j}^{0}} \: h \: \chi_{i}^{0} \: \chi_{j}^{0} \ ,
\end{eqnarray}
 with 
\begin{eqnarray}
 & Y^{h\chi^{-} \chi^{+}}   =  \sqrt{2}\: \xi_{12} \: \frac{\omega}{v}, \nonumber \\
 & Y^{h\chi_{1}^{0} \chi_{1}^{0}}  = \frac{\sqrt{2} \, \omega}{v} \: (\xi_{12} + y - y_{12} ), \nonumber \\
 & Y^{h\chi_{2}^{0} \chi_{2}^{0}}  = \frac{\sqrt{2}\, \omega}{v} \: (\xi_{12} - y - y_{12} ), \nonumber \\
 & Y^{h\chi_{1}^{0} \chi_{2}^{0}}  = 0. \label{Yuk_coup} 
\end{eqnarray}

Interestingly, due to the c.c. symmetry, the interaction of the WIMP ($\chi_{1}^{0}$) with the Higgs follows
$Y^{h\chi_{1}^{0} \chi_{1}^{0}} \sim \xi_{12} + y - y_{12}$. Thus current Direct Detection experimental constraints (discussed later) can be avoided easily without the need for parameter fine tuning. 

Since $D_{1,2}$ are charged under $SU(2)_{L} \times U(1)_Y$, there are renormalizable interactions between them and the corresponding 
gauge bosons. The neutral ones are given by

\begin{eqnarray}
  & \mathcal{L}^{\rm dim=4}_{neutral} \supset  - 
  (+e) \:(\chi^{+})^{\dagger} \bar{\sigma}^{\mu}
  \chi^{+} \: A_{\mu} - (-e)\: (\chi^{-})^{\dagger} \bar{\sigma}^{\mu}
  \chi^{-} \: A_{\mu} \;+ \nonumber \\ 
  & \frac{g}{c_{W}} O^{\prime \, L} \: (\chi^{+})^{\dagger}
  \: \bar{\sigma}^{\mu}\: \chi^{+} \:Z_{\mu} -
  \frac{g}{c_{W}} O^{\prime \, R} \: (\chi^{-})^{\dagger}
  \: \bar{\sigma}^{\mu}\: \chi^{-} \:Z_{\mu}  + \nonumber \\
  & \frac{g}{c_{W}} O_{ij}^{\prime\prime \, L} \: (\chi_{i}^{0})^{\dagger}
  \: \bar{\sigma}^{\mu}\: \chi_{j}^{0} \:Z_{\mu} , 
  \label{L_dim-4_neut}
\end{eqnarray}
where
\begin{equation}
O^{\prime\, L} = O^{\prime \, R} = -\frac{1}{2}(1-2 s_{W}^{2}) \,\,\, and \,\,\,
O^{\prime\prime\, L} = -\frac{i}{2}\left(\begin{array}{cc}0 & 1 \\-1 & 0\end{array}\right),
\label{OLR}
\end{equation} 
with $s_{W} = sin \theta_W$ and $\theta_W$ being the weak mixing angle. Notably, the interaction $\chi_{0}^{1}\chi_{0}^{1}Z$ vanishes due to the charge conjugation symmetry.   

Furthermore, the non-renormalizable operators of eq.~\ref{dipoles} also contribute to the neutral interactions, where the 3-point ones are:

\begin{eqnarray}
 \mathcal{L}_{\rm neutral \; 3-point}^{\rm dim=5} \supset
  &- \frac{\omega}{v^2}\; (d_{\gamma}\; s_W \;+ \;d_W \;c_W) \;  O^{\prime \prime L}_{ij} \: \chi_{i}^{0} \: \sigma_{\mu\nu} \: \chi_{j}^{0} \: F_{Z}^{\mu\nu}- \nonumber \\ 
  & \frac{\omega}{v^2}\; (d_{\gamma} \;s_W \; - \; d_W \;c_W) \: \chi^{-} \: \sigma_{\mu\nu} \: \chi^{+} \: F_{Z}^{\mu\nu} + \nonumber\\ 
  & \frac{\omega}{v^2}\; (d_{\gamma} \;c_W \; - \; d_W \;s_W) \;  O^{\prime \prime L}_{ij} \: \chi_{i}^{0} \: \sigma_{\mu\nu} \: \chi_{j}^{0} \: F_{\gamma}^{\mu\nu} + \label{3-point_neutral} \\
  & \frac{\omega}{v^2}\; (d_{\gamma} \;c_W \; + \; d_W \;s_W) \: \chi^{-} \: \sigma_{\mu\nu} \: \chi^{+} \: F_{\gamma}^{\mu\nu}+
    \ H.c.,\nonumber
\end{eqnarray}
where $F_{\gamma}$ and $F_{Z}$ are the field strength tensors of the photon and $Z$, respectively.
Interestingly, the  dipole operators which arise at $d=5$ level, generate interactions between the neutral dark particles and the photon 
proportional to $C_{\gamma}\equiv  d_W \;s_W \; - \;d_{\gamma} \;c_W$.
There are also other interactions between $W^{\pm}$ and the Dark Sector. The renormalizable ones are:

\begin{eqnarray}
   \mathcal{L}_{\rm charged \; 3-point}^{\rm dim=4} \supset & g\: O_{i}^{L} \: (\chi_{i}^{0})^{\dagger} \: \bar{\sigma}^{\mu} \: \chi^{+} \: W_{\mu}^{-}
   -g \: O_{i}^{R} \: (\chi^{-})^{\dagger} \: \bar{\sigma}^{\mu} \: \chi^{0}_{i} \: W_{\mu}^{-}+ \nonumber\\ 
   &  g \: O_{i}^{L*} \: (\chi^{+})^{\dagger} \: \bar{\sigma}^{\mu}\: \chi^{0}_{i}\: W_{\mu}^{+} 
   -g\: O_{i}^{R*} \: (\chi_{i}^{0})^{\dagger} \: \bar{\sigma}^{\mu}\: \chi^{-} \: W_{\mu}^{+},\label{L_dim-4_charged}
 \end{eqnarray}
with
   \begin{eqnarray}
    O_{i}^{L} =  \frac{1}{2} \left(\begin{array}{c}i \\-1\end{array}\right) \;, 
    O_{i}^{R} =  \frac{1}{2} \left(\begin{array}{c}i \\-1\end{array}\right)  \;.
  \end{eqnarray}  
From eq.~\ref{dipoles}, the 3-point interactions between $W^{\pm}$ and the dark fermions become   
 \begin{eqnarray}
   \mathcal{L}_{\rm charged \; 3-point}^{\rm dim=5} \supset
   &  -2\frac{\omega}{v^2}\;  d_W \;  O^{R\, *}_{i} \: \chi^{-} \: \sigma_{\mu\nu} \: \chi_{i}^{0} \:  F_{W^{+}}^{\mu\nu}  + \nonumber \\
   & 2\frac{\omega}{v^2}\;  d_W \;  O^{L}_{i} \: \chi^{+} \: \sigma_{\mu\nu} \: \chi_{i}^{0} \:  F_{W^{-}}^{\mu\nu} \ + \  H.c.\label{3-point_charged}    
 \end{eqnarray}

Finally, we should point out that, due to an alignment of couplings in eqs.~\ref{L_dim-4_neut} and~\ref{L_dim-4_charged}  with
the those in eqs.~\ref{3-point_neutral} and~\ref{3-point_charged}, a ``natural" cancellation of the d=4 and $d=5$ contributions in the annihilation cross-section  $\chi_{1}^{0}\chi_{1}^{0} \to V V$ (with V being W and Z) can be achieved. This, as we shall see later, will be important in obtaining the observed relic abundance for WIMP masses at the electroweak scale. 

\section{``Earth constraints"}
\label{sec:Earth}
In this section we  study constraints, from WIMP($\chi_1^0$)-nucleon scattering experiments, searches for heavy charged fermions at LEP and from the LHC data for the Higgs boson decay to two photons. We collectively refer to these as ``Earth constraints".

\subsection*{Nucleon-WIMP direct detection bounds}

For the Spin-Independent cross-section the current limit set by the LUX Collaboration
~\cite{Akerib:2015rjg,Akerib:2013tjd} is  $\sigma_{\rm SI} \sim \{1 - 3.5 \}   10^{-45} ~\mathrm{cm}^{2}$, for $ m_{\mathrm {DM}} \sim \{100 -  500 \} \mathrm{GeV}$. 
This translates to

\begin{equation}
|Y^{h\chi_{1}^{0} \chi_{1}^{0}}| \ \lesssim \ \{0.04,0.06\}\,.
\label{dircon}
\end{equation}

\subsection*{LEP bounds}

We next examine constrains for heavy charged fermions from LEP. From Fig.~\ref{hierarchies} we observe  the next-to-lightest particle is the charged dark fermion, $\chi^\pm$,  with mass $m_{\chi^{\pm}}=M_D + \xi_{12} \, \omega$ that is assumed to be positive. 

The bound on $m_{\chi^\pm}$ from such experiments is~\cite{LEP:2001qw}  $m_{\chi^{\pm}}  \ \gtrsim \ 100  \; \mathrm{GeV}$,
which in terms of $\xi_{12}$, $\omega$ and $M_D$ becomes $\xi_{12} \ \gtrsim  \ \frac{100-M_{D}}{\omega}$.

\subsection*{Bound from $h\to \gamma\gamma$ measurements}

From the interactions~\ref{L_Yuk} and~\ref{3-point_neutral},  the ratio $R_{h\to \gamma\gamma}
\equiv \frac{\Gamma(h\to \gamma\gamma)}{\Gamma(h\to \gamma\gamma)_{\mathrm{(SM)}}}$ 
is given by~\cite{Dedes:2014hga}
\begin{equation}
R_{h\to \gamma\gamma} \ = \  
\: \biggl |\: 1 \ + \  
\frac{1}{A_{\mathrm{SM}}} \:   
 \frac{\sqrt{2}\: Y^{h\chi^{-}\chi^{+}}\, v}{m_{\chi^{\pm}}} \: A_{1/2}(\tau)\: \biggr |^{2}\;,
\label{hi2g}
\end{equation}
where $A_{\mathrm{SM}} \simeq -6.5$ for $m_{h}=125$ GeV, $\tau \equiv m_{h}^{2}/4 m_{\chi^{\pm}}^{2}$ and $A_{1/2}$ is the well known function given in Ref.~\cite{Djouadi:2005gi}. The ratio $R$  is currently under experimental scrutiny  at LHC. The current  
value is $R_{h\to \gamma\gamma} = 1.15^{+0.28}_{-0.25}$~\cite{Aad:2015zhl}. From eq.~\ref{Yuk_coup} we expect that $\xi_{12}$ would be restricted to small values from the loop induced $h\to \gamma \gamma$ bound. This would also result to a lower bound on $M_{D}$ at $\sim 100~\mathrm{GeV}$.

\subsection*{Combined ``Earth constraints"}

\begin{figure} 
\begin{center}
\includegraphics[width=0.75\linewidth]{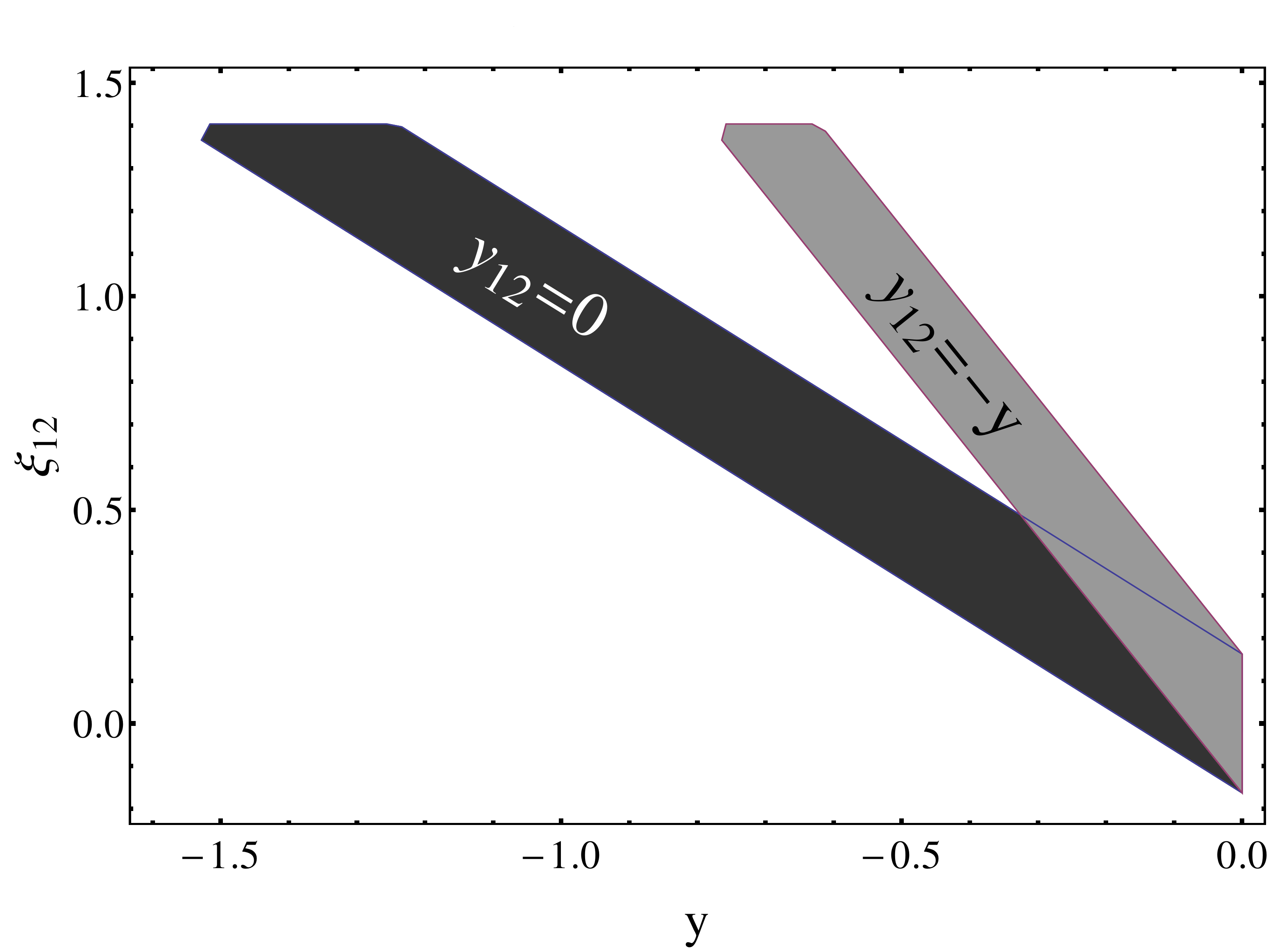}
\end{center}
\caption{The allowed values of $y$--$\xi_{12}$, in order to satisfy the earth constraints, for $\Lambda = 1 \, \mathrm{TeV}$ and $M_D=300 \, \mathrm{GeV}$.}
\label{Combined-Earth}
\end{figure}

A numerical example of the combination of the ``Earth constraints" is shown in Fig.~\ref{Combined-Earth}.
Generally, combining all the aforementioned bounds, results to a lower allowed value for  the doublet mass parameter ($M_D$) at $\sim 90 \, \mathrm{GeV}$. Furthermore, $\xi_{12}$ is restricted to (relatively) small values. Also the allowed Yukawa couplings follow the relation $\xi_{12} \approx - (2) y \pm 0.16$, for $y_{12}=0$ ($y_{12}=-y$). This relation, then, also restricts $y$ to small values.

\section{Cosmological and astrophysical constraints}
\label{sec:astro}
Having examined the constraints imposed from earth-based experiments, we now can the calculate the relic abundance of the lightest particle ($\chi^{0}_{1}$) of this model and delineate the parameter space in which $\Omega h^2$ the observed one. After that, various others astrophysical constraints are going to be considered.  

\subsection*{The role of the dipoles}
Before moving on, we should remind that the dipole operators (\ref{dipoles}) are essential in our study. This is because $d_W$ acts as a regulator 
that minimizes the total annihilation cross-section as the desired EW WIMP mass tends to amplify it\footnote{Generally the cross section scales as 
$M_{D}^{-2}$ at the renormalizable level.}. Thus this minimization  is vital for obtaining cosmologically acceptable relic abundance for 
$m_{\chi^{0}_{1}}$ at the electroweak scale. 

\begin{figure} 
\vspace{-0.cm}\begin{center}
\includegraphics[width=0.75\linewidth]{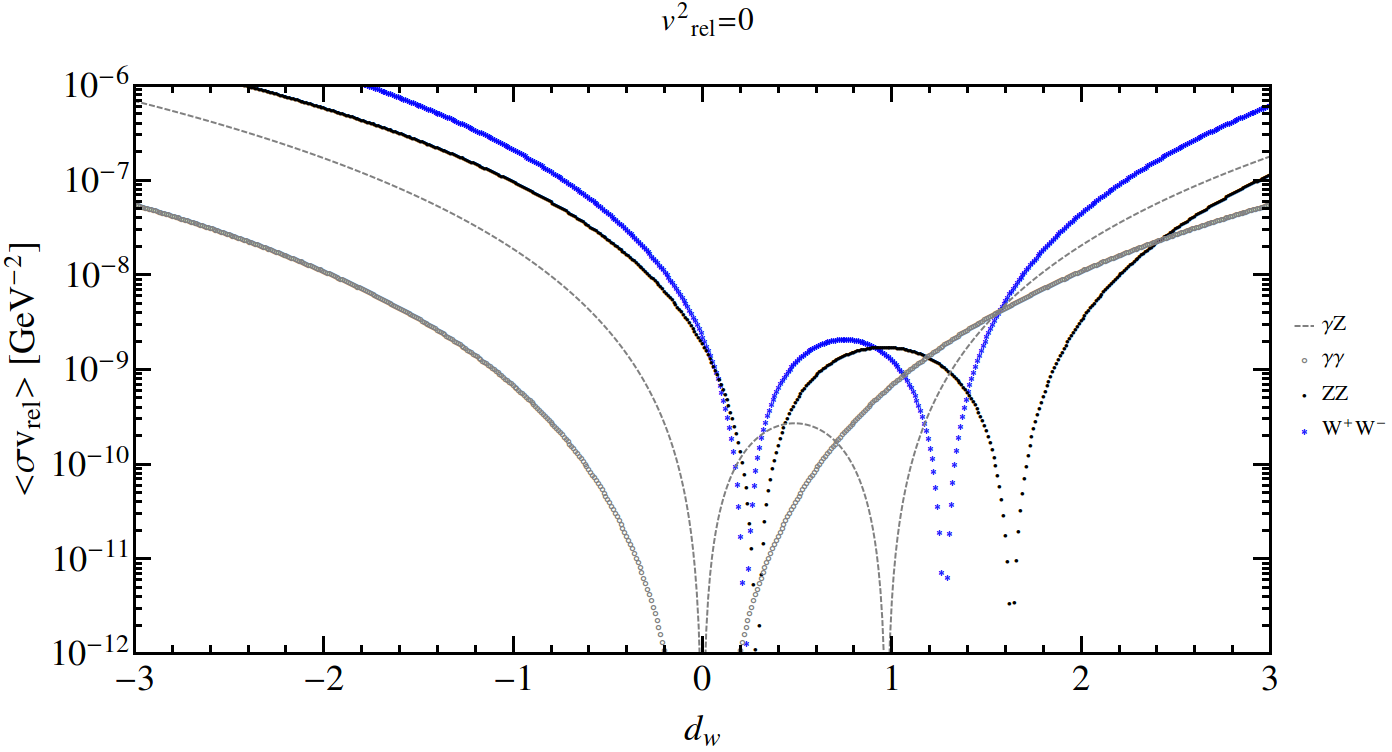}
\end{center}
\caption{The dependence of the cross section for the various annihilation channels on $d_W$ for vanishing relative velocity, $M_D =400  
\mathrm{GeV}$, $\Lambda =1  \mathrm{TeV}$, $y=-y_{12}=-\frac{\xi_{12}}{2}=-0.8$ and $d_{\gamma}=0$.}
\label{csxv-dw}
\end{figure}

The behaviour of the annihilation cross sections of Fig.~\ref{csxv-dw} shows that there are two minima for the channels $\chi^{0}_{1}\chi^{0}_{1} 
\to ZZ$ ($ZZ-$channel) $W^{+}W^{-}$ ($WW-$channel) and $\gamma Z$ ($\gamma Z-$channel) 
and one minimum for $\gamma \gamma$ ($\gamma \gamma-$channel).  The first minimum of the annihilations to $ZZ$ and $W^{+}W^{-}$  coincides with 
$d_{\gamma} \;c_W \; \simeq \; d_W \;s_W$ ($C_{\gamma} \simeq 0$) which gives small cross-sections for the $\gamma \gamma -$  and  $\gamma Z-$ channels.  On the other hand, the second minimum of the $ZZ-$ and $WW-$ channels is in a region where the 
annihilation to $\gamma\gamma$ and $\gamma Z$ blows  up. Furthermore, for negative $d_W$, there are no minima and every cross section becomes quite 
large.  Therefore, if the minimization of the cross section is indeed needed, we expect $d_{W}$ to be bounded to non-vanishing positive values close to $C_{\gamma} \simeq 0$.

\subsection{Relic abundance constraints}
Since the role of the dipole operators is pointed out, we can calculate the relic abundance and set further constraints on the parameter space. 
\begin{figure}
\vspace{-0.cm}\begin{center}
\includegraphics[width=0.52\linewidth]{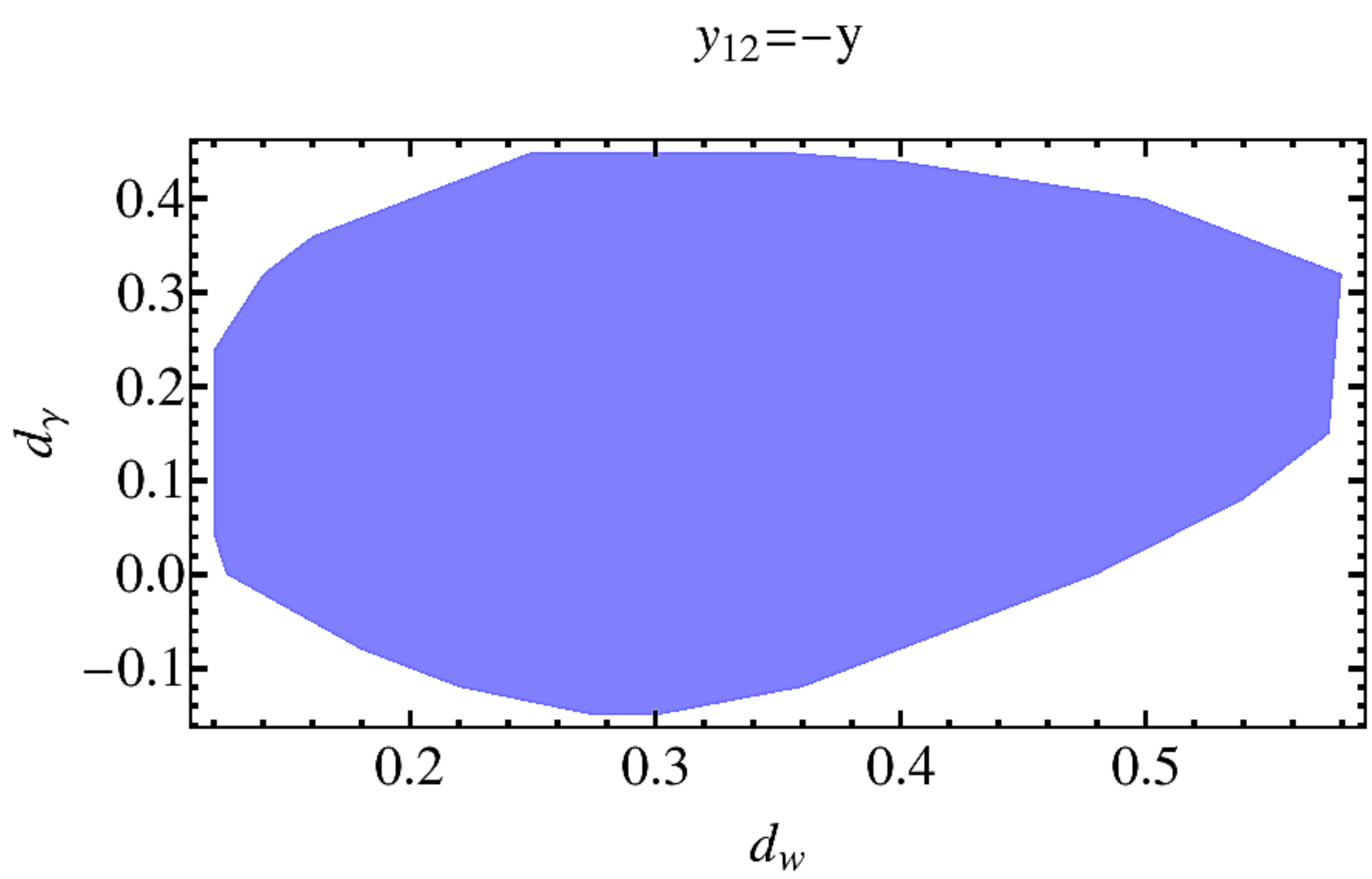}
\caption{The plane $d_W - d_{\gamma}$ of the parameter space that gives the observable relic abundance, for $\Lambda =1 \, \mathrm{TeV}$, $y_{12}=-y$ and allowing the other parameters to vary. Similar region holds for $y_{12}=0$. }
\label{dw-dgamma}
\end{center}
\end{figure}
In Fig.~\ref{dw-dgamma} we show  the  $d_{\gamma} - d_{W}$ plane of the parameter space that is compatible with the observed DM relic density,  varying all the other parameters, while  keeping   $\Lambda =1\,\mathrm{TeV}$ and $M_D \lesssim 500  \mathrm{GeV}$. The parameter $d_{W}$ ($d_{\gamma}$) is bounded to be (mostly) positive in order to explain the DM relic abundance for a WIMP mass at electroweak scale, as expected from the minimization of the annihilation cross section discussed in the previous paragraph.\\

 \begin{figure}
\hspace*{-0.3cm} \begin{subfigure}[b]{0.52\textwidth}
        \includegraphics[width=\textwidth]{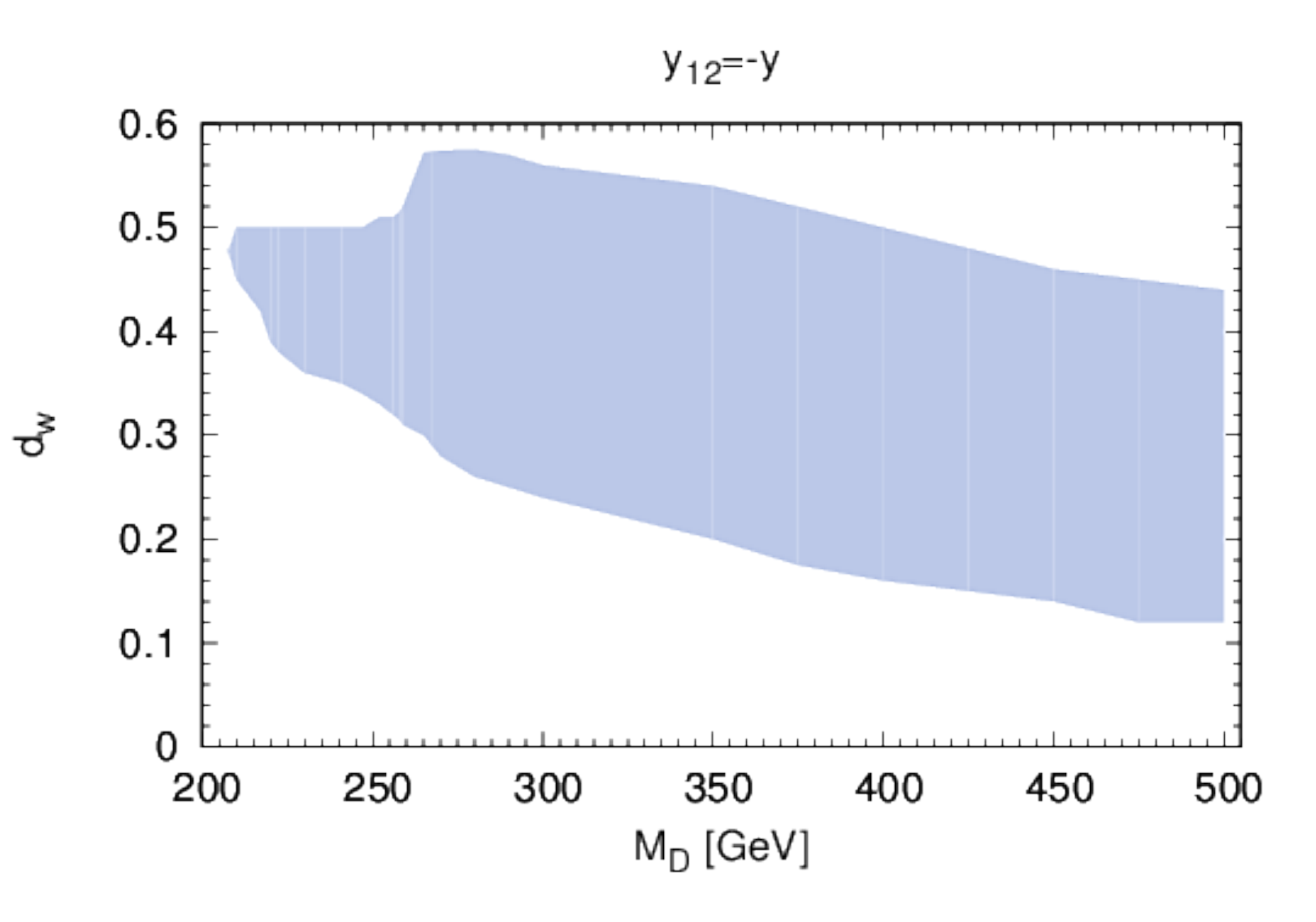}
    \end{subfigure}
\begin{subfigure}[b]{0.52\textwidth}
        \includegraphics[width=\textwidth]{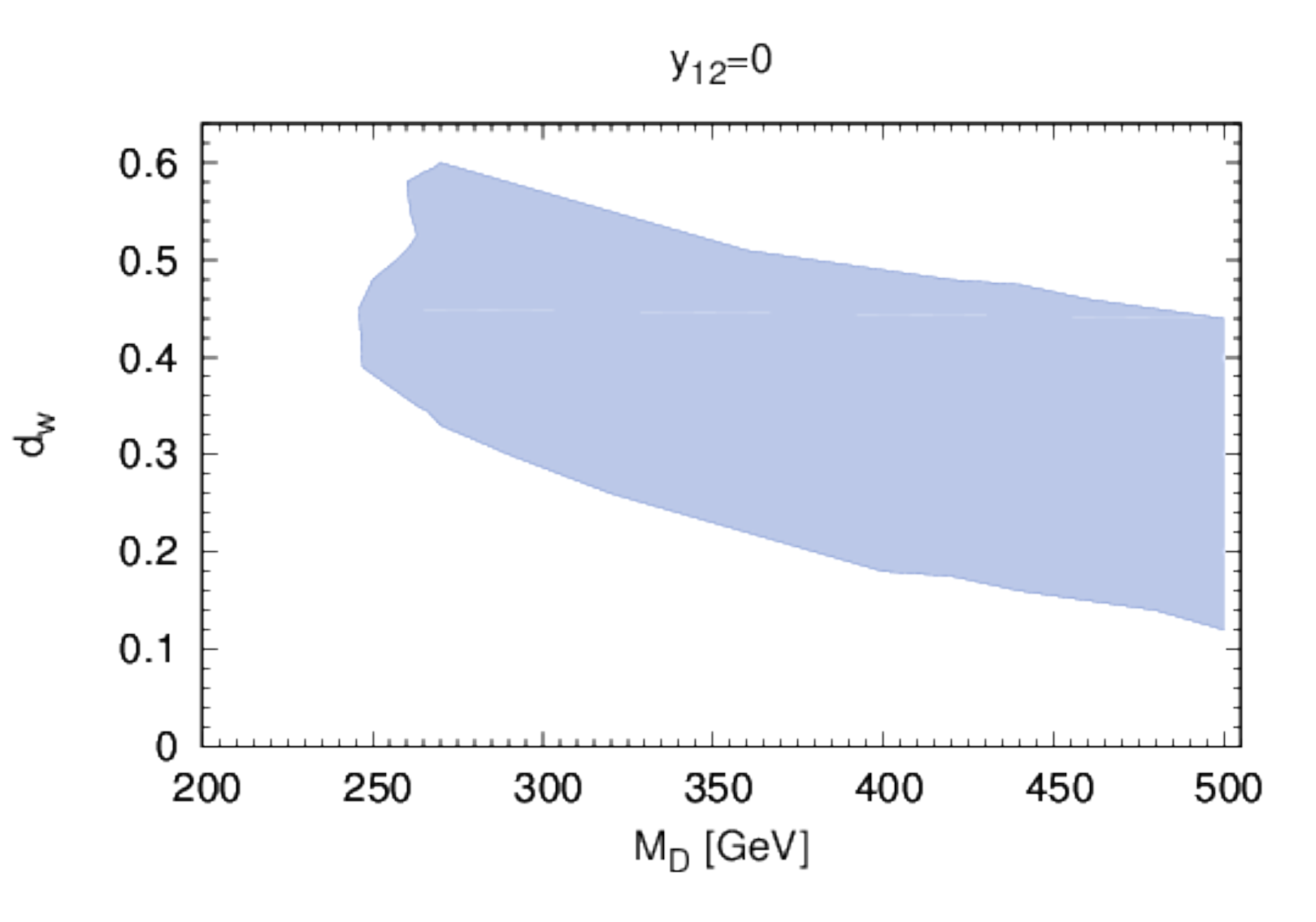}
    \end{subfigure}
\caption{The acceptable values on the plane $M_{D} - d_{W}$ for the two cases (a) $y_{12}=-y$ and (b) $y_{12}=0$,  for $\Lambda =1 \, \mathrm{TeV}$. Again we allow for the other parameters to vary.}
\label{MDvsdw}    
\end{figure}
 \begin{figure}
\hspace*{-0.3cm} \begin{subfigure}[b]{0.52\textwidth}
        \includegraphics[width=\textwidth]{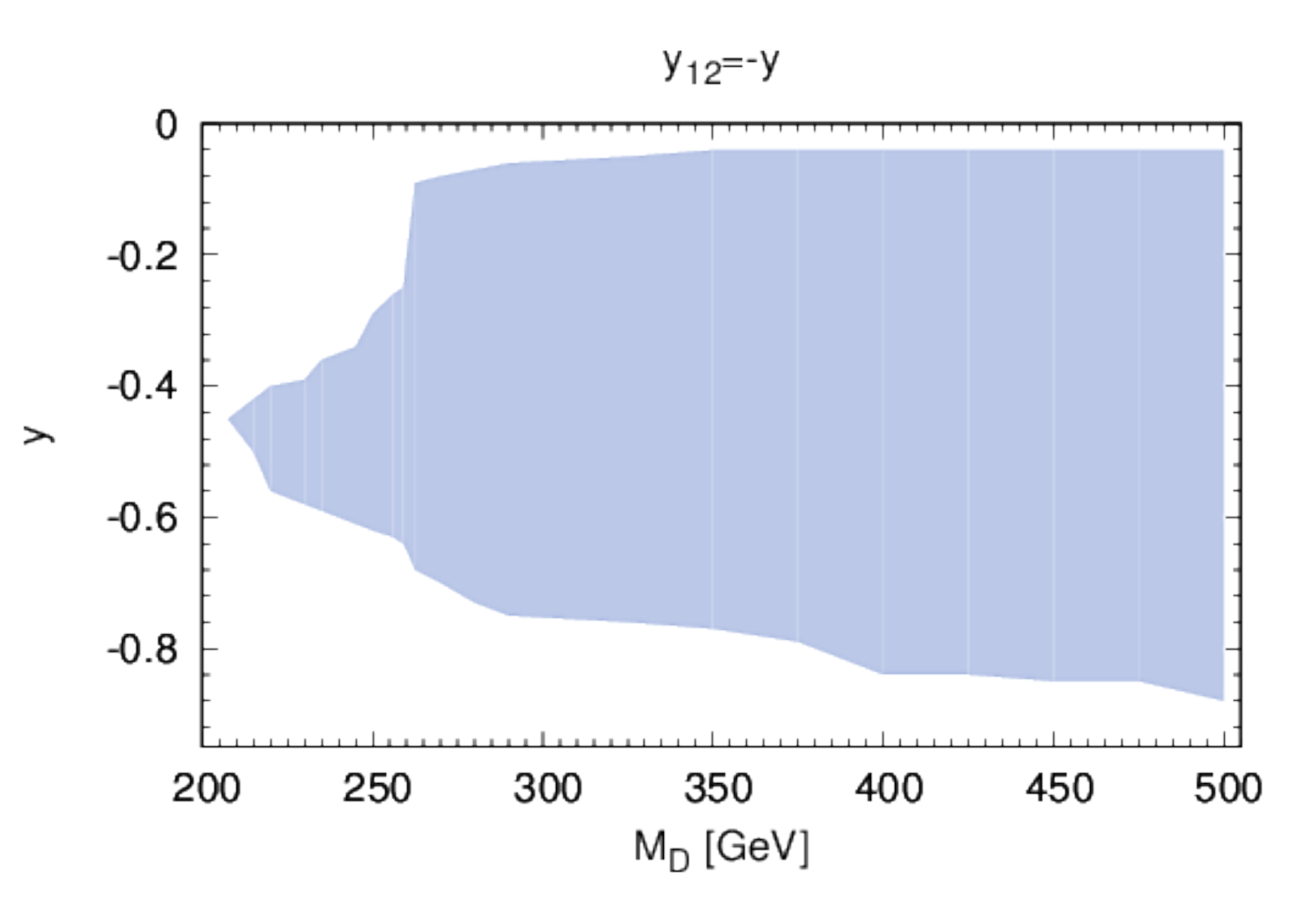}
        \label{y12=-y_MDvsy1_1TeV}
    \end{subfigure}
 \begin{subfigure}[b]{0.52\textwidth}
        \includegraphics[width=\textwidth]{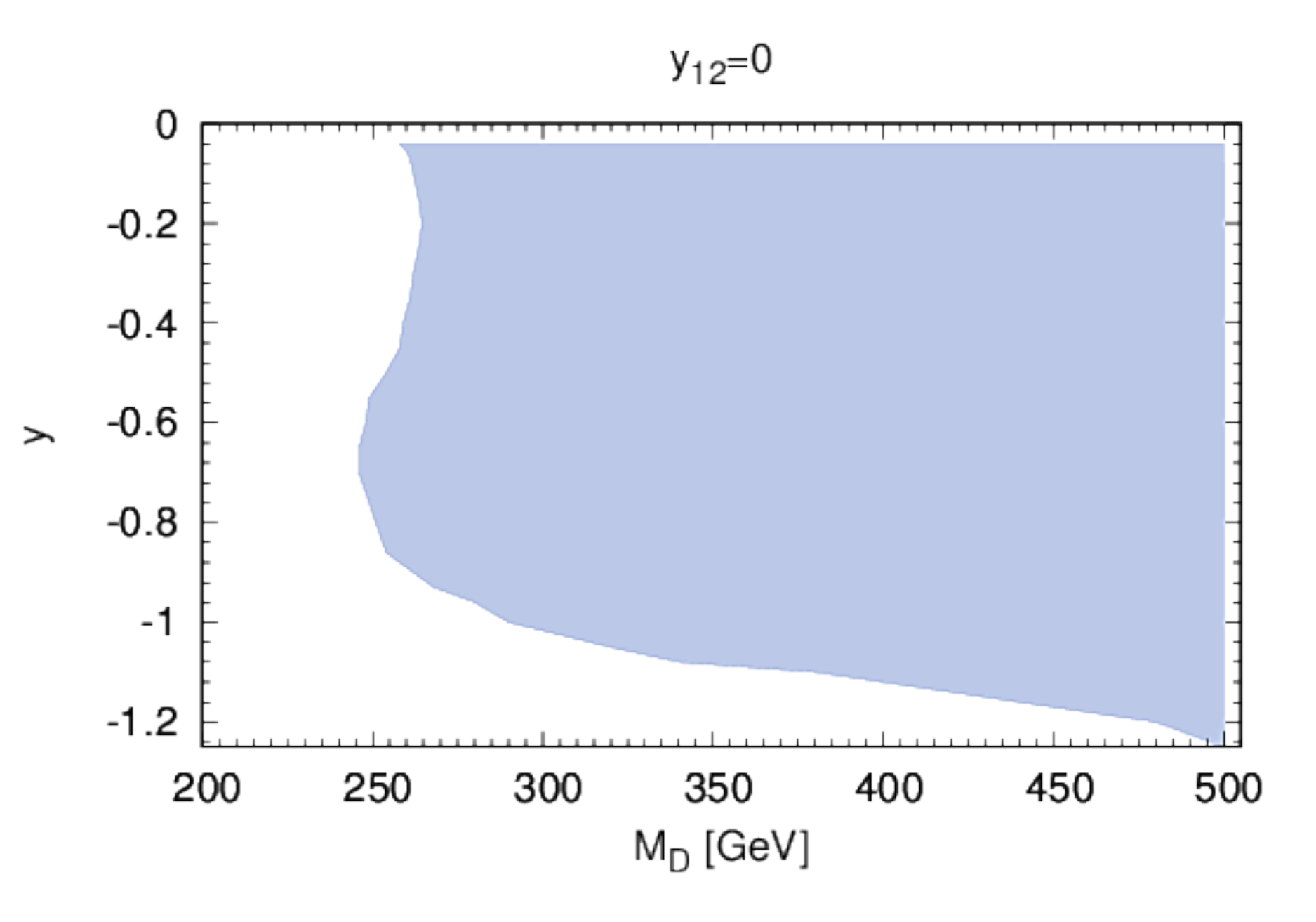}
        \label{y12=0_MDvsy1_1TeV}
    \end{subfigure}
\caption{The acceptable values on the plane $M_{D} - y$ for the two cases (a) $y_{12}=-y$ and (b) $y_{12}=0$,  for $\Lambda =1 \, \mathrm{TeV}$. The other free parameters are allowed to vary. }
 \label{MDvsy1}
\end{figure}

From Fig.~\ref{MDvsdw} we observe that  $M_D$ vastly  affects the  allowed values for $d_W$ for which we obtain the observed relic abundance. This is mainly due to the dependence of the minimum of the total annihilation cross section on $M_D$. Additionally, as $M_D$ increases, $d_{W}$  moves to lower values, since for larger WIMP masses the minimization of the cross section is less needed (we can obtain the desired relic abundance  at the  renormalizable level). Also the allowed values of $M_{D}-y$ are shown in Fig.~\ref{MDvsy1}.\\ \\ 

\begin{figure}
\centering  \includegraphics[width=0.52\textwidth]{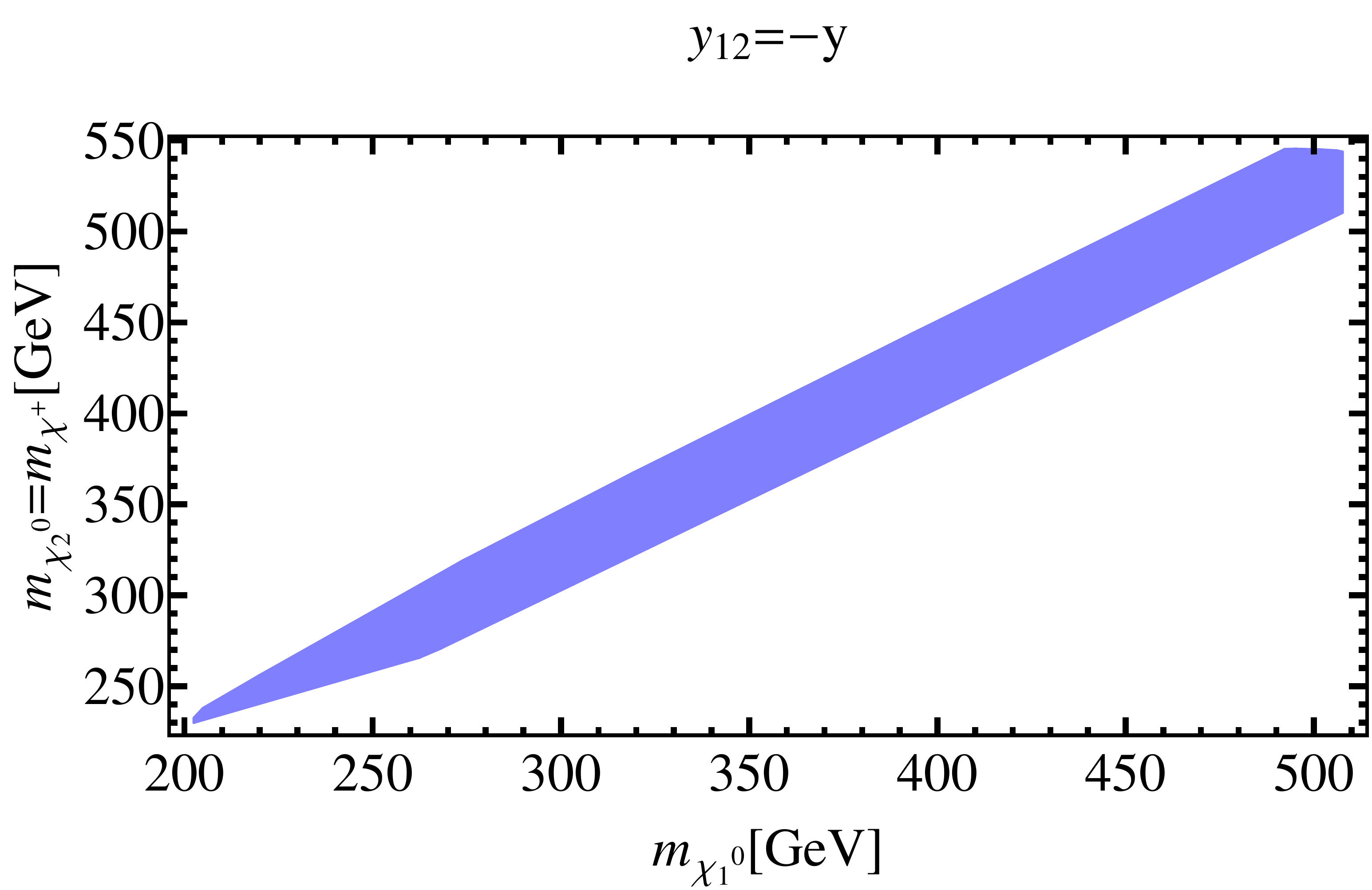}
\caption{The allowed mass region for the case $y_{12}=-y$. A similar region is also allowed for $y_{12}=0$.  }
\label{mass-region}
\end{figure}
Finally, the Yukawa couplings and the mass parameter $M_D$ showed above, fix the masses and their differences, as expected from eq.~\ref{masses}. These masses are shown in Fig.~\ref{mass-region} for $y_{12}=-y$  (similar region holds also for $y_{12}=0$). We observe that the WIMP can be at the EW scale and its mass, as expected, is dominated by $M_{D}$.

\subsection{Constraints from Gamma-ray monochromatic spectrum}
\label{gamma-rays}

In this paragraph, we calculate the cross-sections for processes that could give gamma-ray lines from the Galactic Center (GC). 
As input, we use the  parameter space  that evade all the other, previously examined, restrictions and use the results from Fermi-LAT~\cite{Ackermann:2015lka} to set additional bounds  to the parameters of this model.

\begin{figure}
\hspace*{-0.5cm} \begin{subfigure}[b]{0.53\textwidth}
        \includegraphics[width=\textwidth]{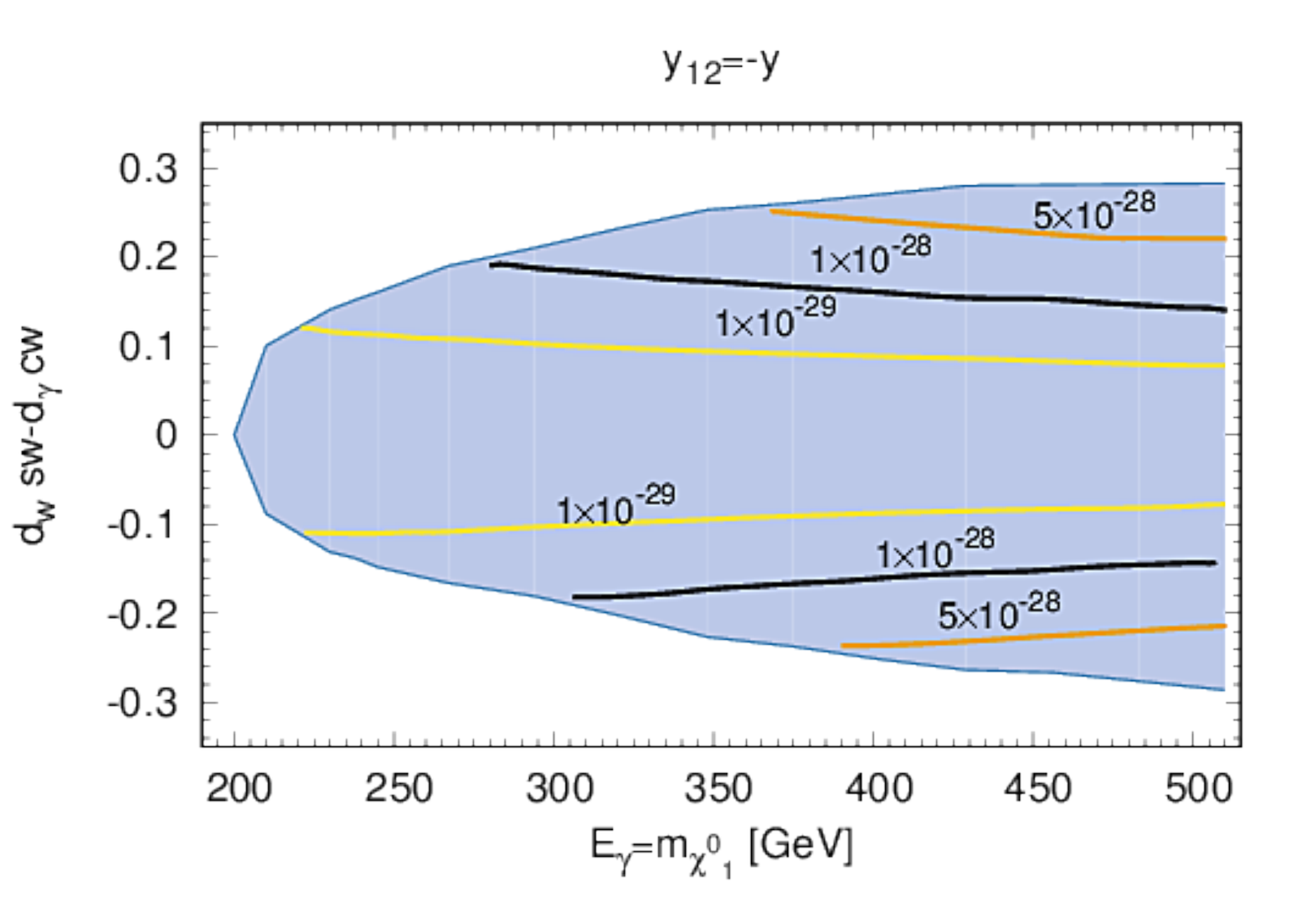}
           \caption{}
    \end{subfigure}
\hspace*{0.0cm} \begin{subfigure}[b]{0.53\textwidth}
        \includegraphics[width=\textwidth]{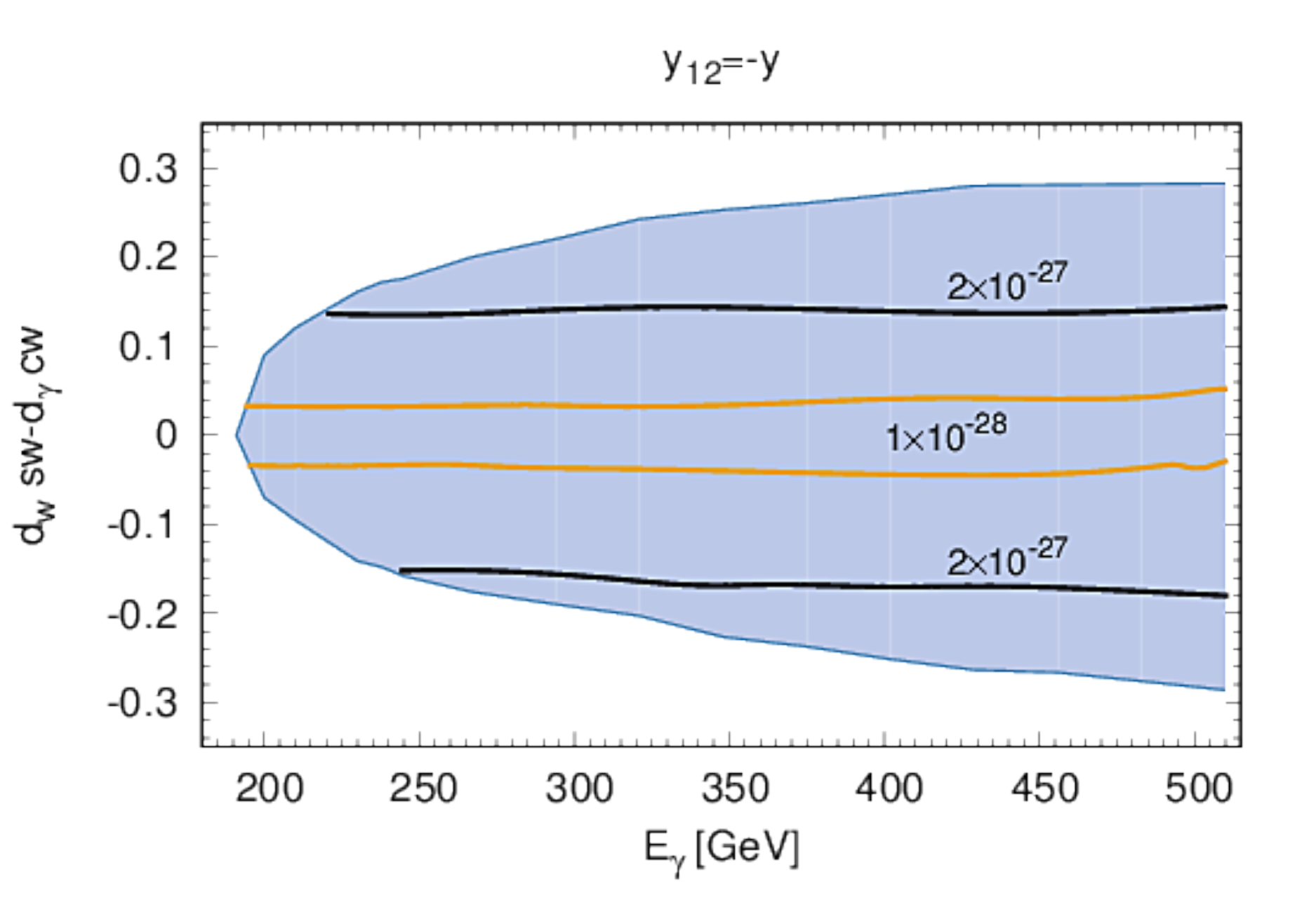}
           \caption{}
    \end{subfigure}
\caption{The allowed,   region of the parameter space, in terms of the photon energy and the coupling $d_{W}\,s_{W}-d_{\gamma}\,c_{W}$. The contours show the values of the  cross-section of the channels  $\chi_{1}^{0}\chi_{1}^{0} \to \gamma \gamma $ (a) and $\gamma Z$ (b) in $cm^{3}s^{-1}$ for $y_{12}=-y$. Again $y_{12}=0$ results in  an almost identical plot.}
        \label{gammagamma-gammaZ-gamma_rays_lines}
\end{figure}
Fro observations of gamma-ray lines from the GC show that the annihilation cross-section for $\chi_{1}^{0}\chi_{1}^{0}\to \gamma \gamma$ cannot be
above   $\sim 10^{-28} \; cm^{3}s^{-1}$ for photon energy ($E_{\gamma}=m_{\chi_{1}^{0}}$) at  $200 \; \mathrm{GeV}$ up to $\sim 5 \times 10^{-28}$ 
for $E_{\gamma}\sim 500 \;\mathrm{GeV}$. For the photon production channel $\chi_{1}^{0}\chi_{1}^{0}\to \gamma Z$, we need to rescale this bound by a factor of two. This process also results to different value for the photon energy given by  $E_{\gamma}=m_{\chi_{1}^{0}}\, (1-  {m_{Z}^2}/{4m_{\chi_{1}^{0}}^2} )$.

The values of the relevant cross sections in the allowed region of the parameter space are shown in Fig.~\ref{gammagamma-gammaZ-gamma_rays_lines}. 
Applying the bounds discussed above, the parameter space remains virtually unaffected, apart from the $d_{W}-d_{\gamma}$ plane shown in 
figure~\ref{dw-dgamma}.

\begin{figure}
\centering
\includegraphics[width=0.52\textwidth]{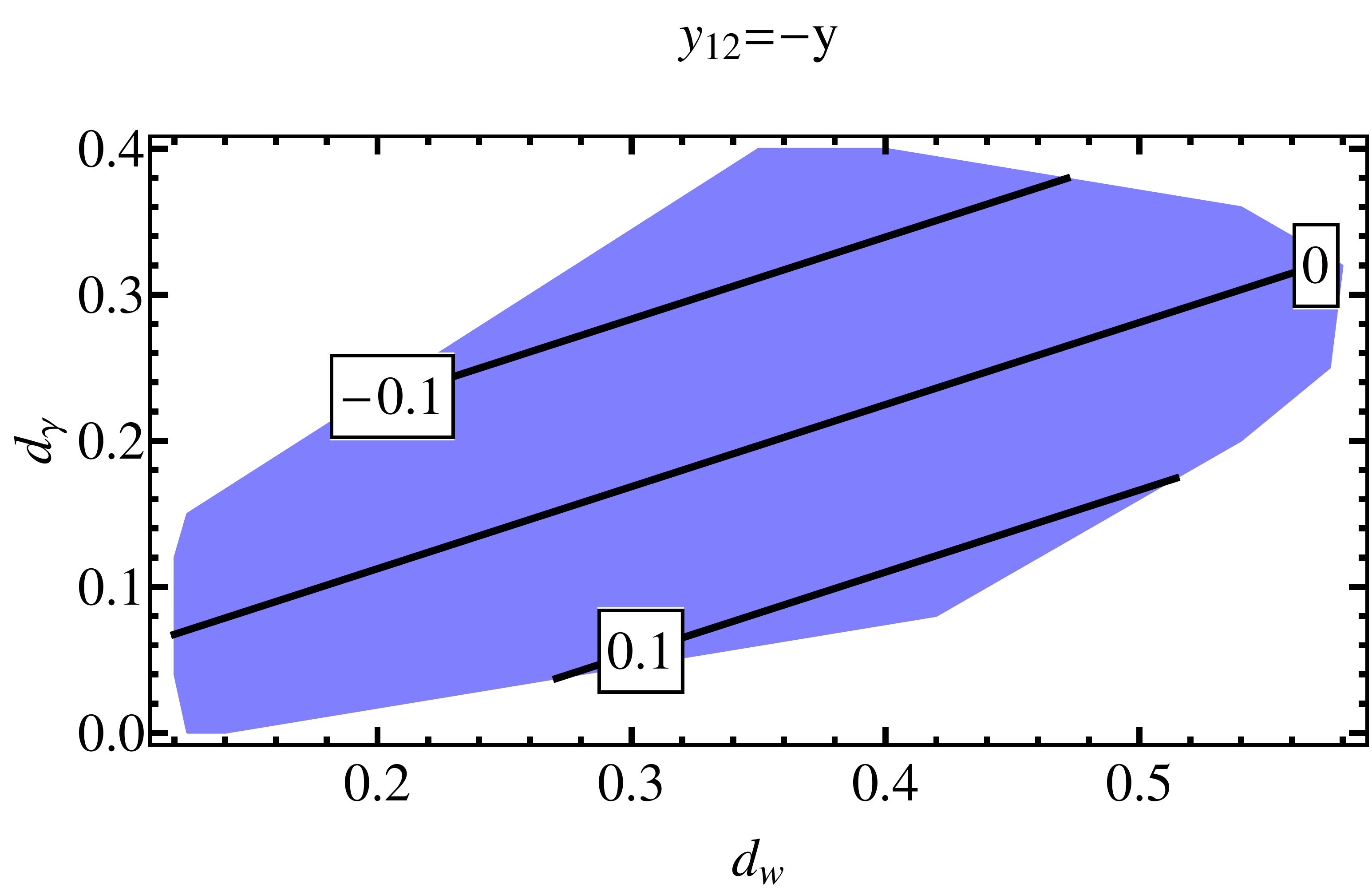}
           \caption{ Allowed  regions on the $M_D - C_{\gamma}$ plane, consistent with ``Earth" constraints, the observed  relic abundance and the bounds from gamma-ray monochromatic spectrum.  Almost identical regions  are allowed  for $y_{12}=0$. 
The contour lines show the values of the $\chi_{1}^{0}\chi_{2}^{0}$-photon coupling $C_{\gamma}$.}
         \label{dwvsdgamma_update}  
\end{figure}

The values of $d_{W}-d_{\gamma}$ which respect the bounds discussed here are shown in Fig.~\ref{dwvsdgamma_update} along with  $C_{\gamma}$ (contours). We observe that the allowed values of $C_{\gamma}$ are concentrated around zero, which  forces $d_{W}$ and $d_{\gamma}$ to have the same sign. Thus the latter is restricted to positive values, while accepted range of $d_W$ remains as in Fig.~\ref{dw-dgamma}.

\section{LHC searches}
\label{LHC}

Having found the viable area in the parameter space, in which the observed DM relic abundance is obtained while avoiding all the other experimental and observational constraints, we move on to discuss possible observational effects at the LHC. 

\begin{figure}
\hspace*{-0.5cm} \begin{subfigure}[b]{0.52\textwidth}
        \includegraphics[width=  \textwidth]{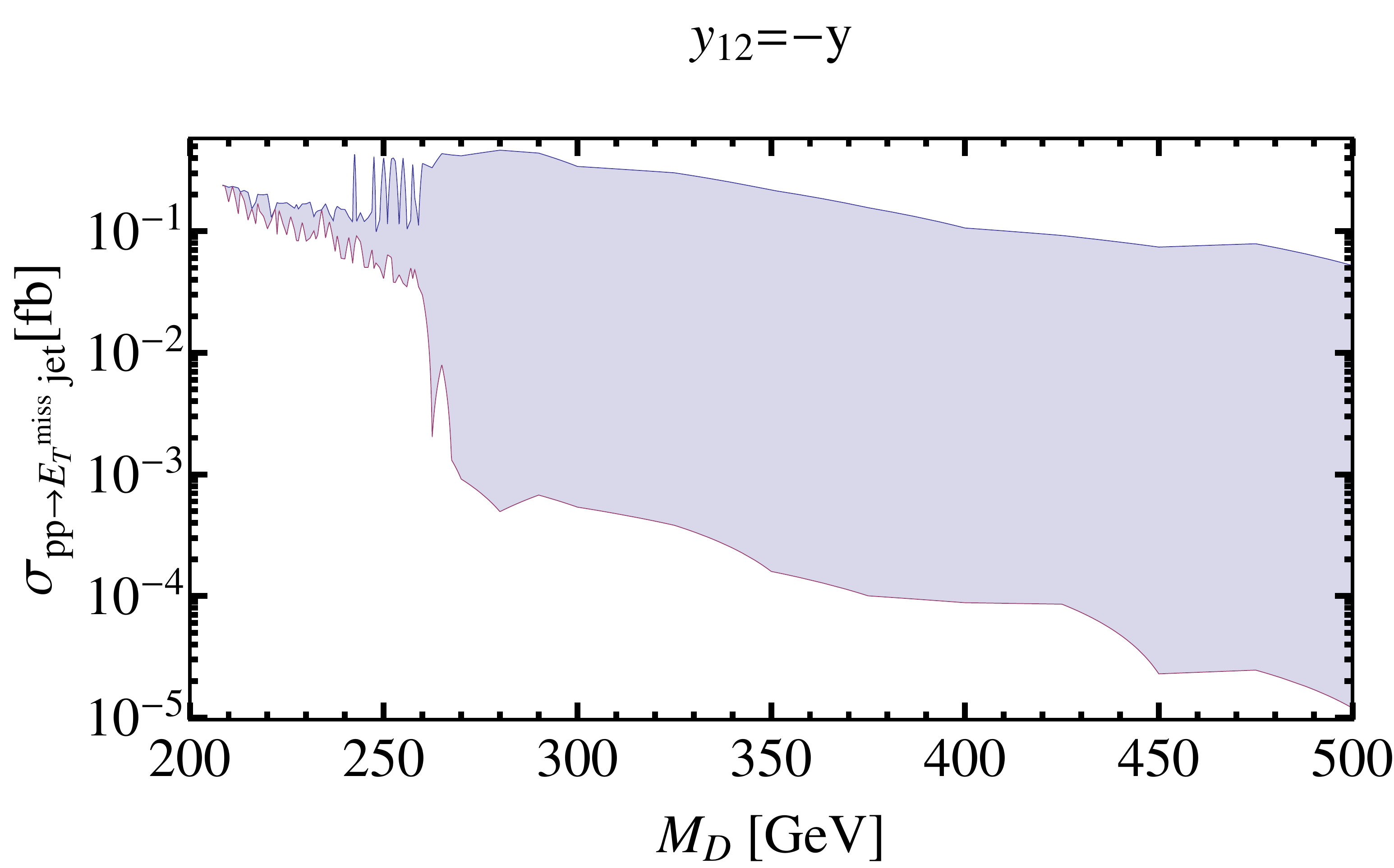}
           \caption{}
         \label{mono-jet_8TeV}
    \end{subfigure}
\hspace*{-0.0cm} \begin{subfigure}[b]{0.52\textwidth}
        \includegraphics[width=  \textwidth]{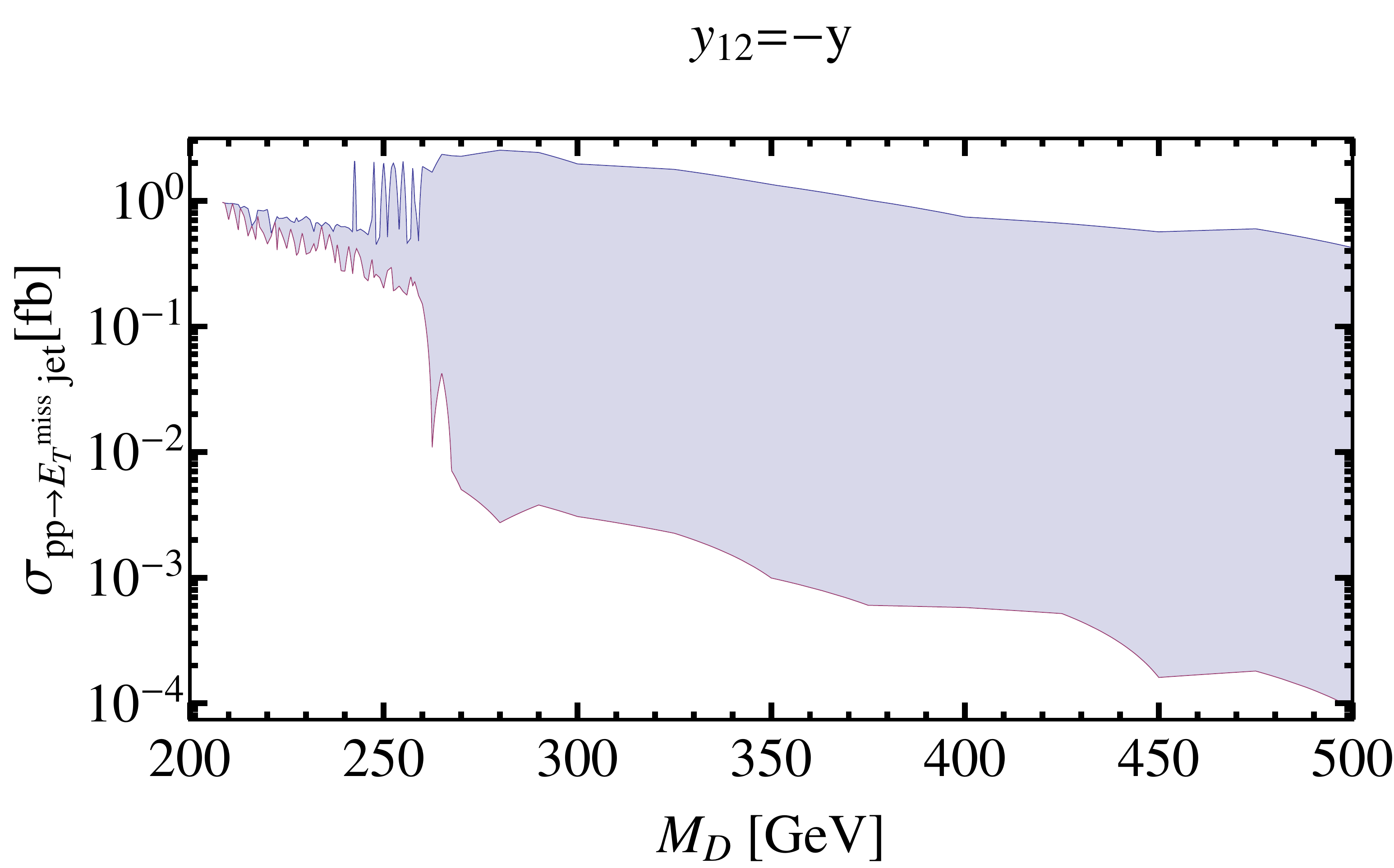}
           \caption{}
         \label{mono-jet_13TeV}
    \end{subfigure}
\caption{The cross-section for the mono-jet channel with (a) $\sqrt{\hat{s}}=8 \mathrm{TeV}$ and (b)  $\sqrt{\hat{s}}=13  \mathrm{TeV}$. The areas are obtained from randomly selected values from the ones satisfying all the constraints discussed in the previous sections. }
        \label{mono-jets}
\end{figure}

In Fig.~\ref{mono-jets} we show the mono-jet channel cross section, which seems to be the most promising one, at least for this model.
The current bound~\cite{Khachatryan:2014rra} on $pp \to   \chi^{0}_{1} \, (\chi^{0}_{2} \to \chi^{0}_{1} \,+\, \nu \bar{\nu}) \,+\, jet$ at center of mass energy $\hat{s}=8 \,\, \mathrm{TeV}$ is   $\sigma_{\slashed{E}_T+jet} \lesssim 6.1 \,fb$. It is apparent that this bound is easily evaded in the allowed parameter space. 
For LHC (RunII) at  $\sqrt{\hat{s}}=13\;\mathrm{TeV}$, the mono-jet channel  can provide us with a relatively large number of events.
From Fig.~\ref{mono-jet_13TeV}, we observe that the production 
of a jet accompanied with missing $E_T$, can reach cross sections  up to $\sim 2.5 \, fb$. 
Therefore the number of events that can, in principle, be observed is around $250 \, (750) $ for LHC expected luminosity  reach of $100 \, (300) \; fb^{-1}$.

\section{Conclusions}
We have extended the SM particle spectrum by a fermionic pair of doublets, $D_{1,2}$,  with opposite hypercharges. 
In addition, we have assumed a discrete $Z_2$-symmetry that distinguishes this Dark Sector from the SM 
fields. At the renormalizable level there are a neutral, and a charged Dirac, fermion. 
After EW symmetry breaking, due the presence of $d=5$ operators, the neutral Dirac fermion splits to two Majorana states.
The lightest of them ($\chi_1^0$), which is the WIMP candidate and a heavier neutral state,  $\chi_2^0$.  Moreover, the $d=5$ operators include  
magnetic dipole operators which  are, in principle, generated by a UV-complete model, at the TeV scale. The question we ask here is whether the 
WIMP, with a mass around the EW scale, is compatible to the various experimental and observational data.

In order to reduce fine tuning and further simplify the parameter space, in section~\ref{sec:EFT_cont}, we adopted two scenarios based on a charge conjugation symmetric limit. 
Then, in section~\ref{sec:physical} we showed the mass spectrum of the physical states. 

In section~\ref{sec:Earth},  we performed an analysis based a) on scattering 
WIMP-nucleus scattering experiments, b) LEP searches for
heavy charged fermions, as well as c) on LHC searches for the decay $h\to \gamma\gamma$. 
We found the collective bounds showed in Fig.~\ref{Combined-Earth}.
 
In section~\ref{sec:astro}  we calculated  $\Omega h^2$ for our  WIMP candidate. 
In the presence of non-vanishing $d=5$ dipole interactions, the  WIMP annihilation cross sections
acquire minima, which allow for the WIMP mass to be as low as $200 \, \mathrm{GeV}$. Following this we
also considered constraints based of monochromatic gamma-ray spectrum observations from the Galactic Center. 
These set the final restrictions on the parameter space, which confined the photon dipole coupling ($C_{\gamma}$)
to be $\sim \pm 0.1$.   
 
Since our main goal was to be able to produce a WIMP at the electroweak scale, in order to be accessible at the LHC,
in section~\ref{LHC} we estimated the cross section for producing $\chi_1^0$ in  association with a jet (monojet) with center of mass energy 
$\sqrt{\hat{s}} = 8, 13$ TeV.  Although current bounds are weak, we found that the monojet channel, can produce few hundred of events at  
$\sqrt{\hat{s}} = 13 $ TeV and with $m_{\chi_1^0} \simeq 200 - 350 \; \mathrm{GeV}$ (see  Fig.~\ref{mono-jets}).

\bibliography{EffDM-PoS.bib}{}
\bibliographystyle{JHEP} 

\end{document}